\documentclass[aps,groupedaddress,twocolumn]{revtex4-1}
\usepackage{graphicx}
\usepackage{xcolor}
\begin{document}


\title{15~cm$^{-1}$ to 12000~cm$^{-1}$ Spectral Coverage without Changing Optics:  Diamond Beam Splitter Adaptation of  an FTIR Spectrometer}


\author{Dmitry Strelnikov}
\email[]{dmitry.strelnikov@kit.edu}
\author{Bastian Kern}
\altaffiliation[current address: ]{Max Planck Institute for Solid State Research, Heisenbergstraße 1, 70569 Stuttgart, Germany}
\author{Christoph S\"{u}rgers}
\altaffiliation{Karlsruhe Institute of Technology, Physikalisches Institut, Karlsruhe, Germany}
\author{Manfred M. Kappes}
\affiliation{Karlsruhe Institute of Technology, Division of Physical Chemistry, Karlsruhe, Germany}


\date{\today}
 
\begin{abstract}
In order to facilitate IR absorption measurements of mass-selected ions isolated in cryogenic matrices, we have upgraded an ion beam deposition apparatus encompassing a Bruker IFS66v/S FTIR spectrometer. A synthetic diamond beam splitter without compensator plate and UHV diamond viewports were installed. We have also modified the IR detector chamber to allow measurements with 5 different detectors. As a result we can now obtain FT absorption spectra from 12000~cm$^{-1}$ to 15~cm$^{-1}$ with the same sample held under ultrahigh vacuum conditions, simply by switching between appropriate IR detectors. We demonstrate performance of the upgraded FTIR spectrometer by presenting measurements of matrix isolated fullerene ions and an adhesive tape.
\end{abstract}


\maketitle

\section{Introduction}
Fourier-transform infrared (FTIR) spectroscopy is a powerful tool to obtain IR spectra over a broad spectral range. The factors limiting spectral coverage are the beam splitter, optional windows and the IR detectors associated with the spectrometer. Each of these parts is designed for a specific wavelength range. Therefore, to change a measurement range one has to switch or replace the above mentioned elements. In the case of samples held under vacuum such changes are time consuming and may even require new sample preparation. Changing the beam splitter associated with the Michelson interferometer contained in an FTIR often requires readjustment of the beam path. Some vacuum FTIR spectrometers offer automatic beam splitter changer units, but in practice the optical alignment which is best  for all beam splitters taken together is not the best one for a specific beam splitter. Also, in some cases one needs a measurement of the entire spectrum at once, without switching between various beam splitters \cite{BMSBruker}. In our investigations of matrix-isolated mass-selected molecular ions \cite{Depo2C60,Depo2C60dicat,Depo2C70} we are interested in the optical absorptions from UV to Far-IR. Some samples require several days of accumulation. In this case one would like to use the same sample for recording spectra in all accessible spectral regions. Diamond optics can provide a solution to this problem. Diamond is a unique material with one of the broadest optical transmission ranges from UV to Far-IR. It is however quite expensive in particular in natural form. Consequently, synthetic diamond beam splitters have been tested for use in FTIR spectrometers \cite{DiamondBSFTIR1}. A prototype FTIR spectrometer CIRS-Lite for space missions\cite{CIRS-Lite,CIRS-Lite1}, equipped with a synthetic diamond beam splitter, was developed by NASA. Recently, enhanced production processes have significantly lowered the price of synthetic diamond \cite{CVDproduction}. As a result, we decided to upgrade our Bruker IFS66v/S spectrometer with diamond optics (diamond beam splitter, diamond windows for a sample chamber) and have modified the detector chamber to allow recording spectra over a wide range from Far-IR to near-IR. The corresponding changes   to the spectrometer are described below together with the resulting  performance enhancements. A similar upgrade can also be applied  to other FTIR spectrometers.

\section{Implementation}
\begin{figure}
\includegraphics[width=0.3\textwidth]{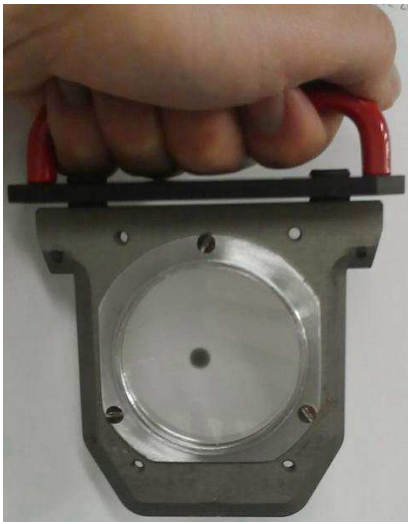}%
 \caption{\label{DBS} Diamond beam splitter with a phase shifting spot.}
 \end{figure}

A synthetic diamond beam splitter (57.1~mm disc diameter, 0.95~mm central thickness, 0.11$^\circ$ wedge, planarity 0.15-0.22$\lambda$, measured with 632~nm HeNe laser) and two synthetic diamond wedged UHV viewports (20~mm free aperture, 0.5~mm central thickness, 0.5$^\circ$ wedge) were obtained from Diamond Materials GmbH (Freiburg, Germany). Wedged optics are needed so as to preclude interference fringes in spectra. An original Bruker holder was used to stably mount the diamond beam splitter. A first attempt to then use the diamond beam splitter without further modifications failed.  
\begin{figure}
 \includegraphics[width=0.5\textwidth]{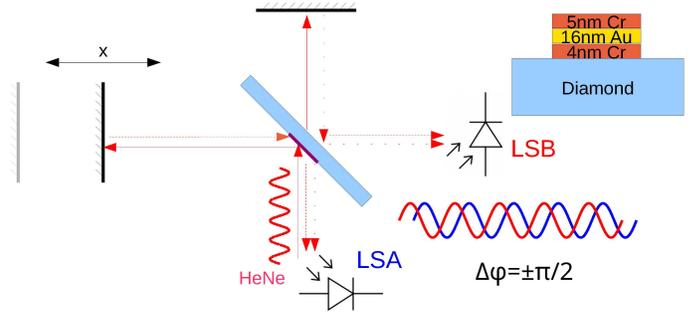}%
 \caption{\label{Michelson} Michelson interferometer with phase quadrature detection of HeNe interference signals.}
 \end{figure}
The reason was that the electronics of the IFS66v/S spectrometer is designed to work with a phase quadrature detection interferometer \cite{Quadrature}. This requires a (90$\pm$30)$^\circ$ phase shift between HeNe laser intensity signals on LSA  and LSB photodiodes at the output and input of the interferometer, as shown schematically in Fig.~\ref{Michelson}. This can be achieved by passing the HeNe laser beam through a custom absorbing metal coating consisting of a ca. 5 mm spot diameter evaporated onto the beam splitter, Fig.~\ref{DBS}. Many different metal coatings were tested on a quartz beam splitter in order to provide the 90$^\circ$ phase shift at 632~nm \cite{Quadrature}. We chose a 3-layer (4~nm Cr, 16~nm Au, 5~nm Cr) coating, because this coating has shown no aging over a six month period\cite{Quadrature}. The coating was prepared in a high-vacuum system (residual pressure 10$^{-6}$~mbar) by magnetron sputtering using Au and Cr targets in 6$\cdot$10$^{-3}$~mbar Ar atmosphere. The deposition rates (0.2 -0.3 nm/s) were calibrated ex situ by measuring the film thicknesses with a profilometer. CVD diamond turns out to be  a very robust material. Hence, one can coat it and remove coatings without surface damage using, for example, soft polishing with cerium oxide powder and wiping the surface with acetone.

 \begin{figure*}
 \includegraphics[width=\textwidth]{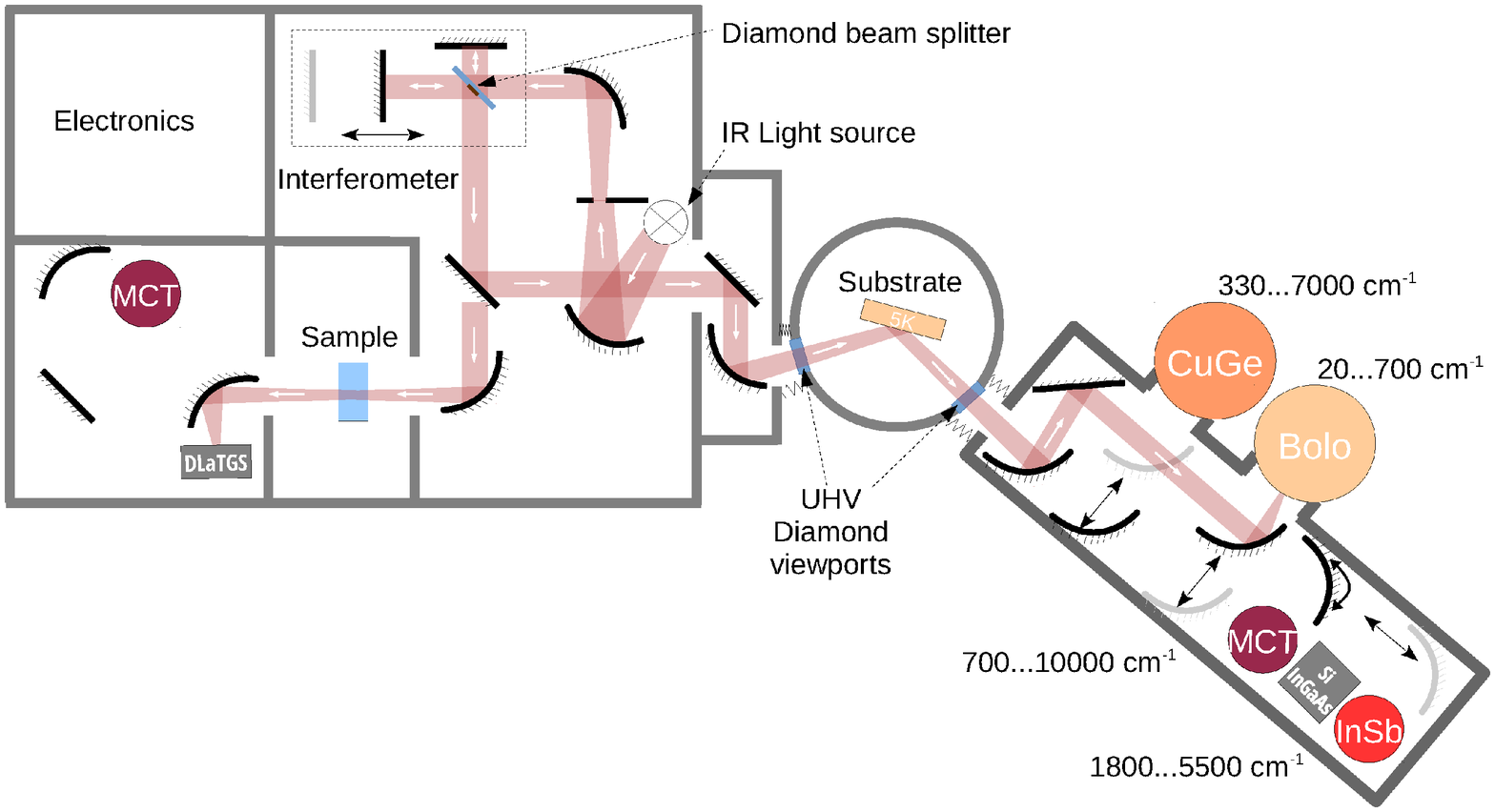}%
 \caption{\label{FTIR} Beam path in the upgraded Bruker IFS66v/S FTIR spectrometer. Detectors on the left side were used just for test purposes.}
 \end{figure*}

To make the interferometer adjustment procedure easier, we replaced the original 1.6~mW HeNe laser by a 10~mW HeNe laser. It turned out, that the performance of the original photodiodes (equipped with 632~nm laser-line filters) was not optimal. Therefore, we replaced them with Hamamatsu S6430-01 red color detectors. The original HeNe laser signal preamplifier (TC20-IFM) was slightly modified: resistors for gain (R5, R10) and offset (R4, R9) were replaced by trim potentiometers (R4, R9 -- 100k; R5, R10 -- 500k). With these modifications we obtained sufficient signal levels at the analog outputs LSA2, LSB2 of the HeNe preamplifier. The previously described coating provided the necessary $\pm$90$^\circ$ phase shift between LSA and LSB signals. From our experience it is easiest to adjust the interferometer by removing the LSB photodiode and then the laser beams otherwise falling onto it up to the ceiling and then superposing them there. After this procedure, the LSB photodiode can be returned to its original position.

We note, that a 45$^\circ$ angle of incidence into the interferometer arm is not the optimal angle in the case of a diamond beam splitter. Fresnell's equations predict that only about 8\% of the white light contributes to the interferogram signal at this angle. Decreasing the angle of incidence would slightly improve the optical throughput, but this would require completely rebuilding the interferometer. We also tried various gold coatings of the entire beam splitter surface (1.5, 3, 6~nm thickness) hoping to get a 50:50 transmission/reflection ratio. It turned out, that the uncoated beam splitter has the best performance. The reason is that the gold layer not only changes the transmission/reflection ratio, it also absorbs the transmitted light. This leads to a decrease of the interferogram intensity.

A Michelson interferometer working with a white light source usually requires a compensator plate, mostly for correction of chromatic aberration. Previous FTIR tests with a diamond beam splitter established that the compensator plate is not really necessary \cite{DiamondBSFTIR1,CIRS-Lite}, because the refractive index of diamond changes insignificantly from near-IR to far-IR. Therefore, we decided to leave a compensator plate out.

 Fig.~\ref{FTIR} shows the external detector chamber together with a schematic of the overall setup (substrate refers to a vacuum chamber which houses the cryogenic matrix typically held at a pressure of 10$^{-9}$--10$^{-8}$~mbar, see \cite{Depo2C60,Depo2C60dicat}). The external IR detector chamber was designed and built in such a way that one can switch between five different IR detectors ((i) CuGe -- Infrared Laboratories 4.2~K CuGe detector; (ii) Bolo -- Infrared Laboratories 4.2~K Bolometer; (iii) MCT -- Electro-Optical Systems Inc. MCT(10)-010-E-LN4 HgCdTe detector; (iv) Si/InGaAs -- Hamamatsu K3413-08 two color detector; (v) InSb -- Judson J10D detector). A DLaTGS -- Deuterated L-alanine doped TriGlycine Sulphate pyroelectric detector mounted in the internal detector chamber was used for test purposes only. The FTIR spectrometer is equipped with a switchable light source unit for two light sources. Gold mirrors were installed on motorized linear translation stages  to allow fully-automatized switching between detectors. A focusing mirror for MCT, Si/InGaAs and InSb detectors was additionally equipped with three tilting stepper motors. The control electronics is based on an Arduino-compatible uC32 Digilent microcontroller. We note, that since  diamond is also transparent in the visible spectral range, the optical adjustment procedure turned out to be quite simple, especially compared to working with IR materials which are only transparent in the infrared range. 

\section{Performance Tests}
\begin{figure}%
\includegraphics[width=0.5\textwidth]{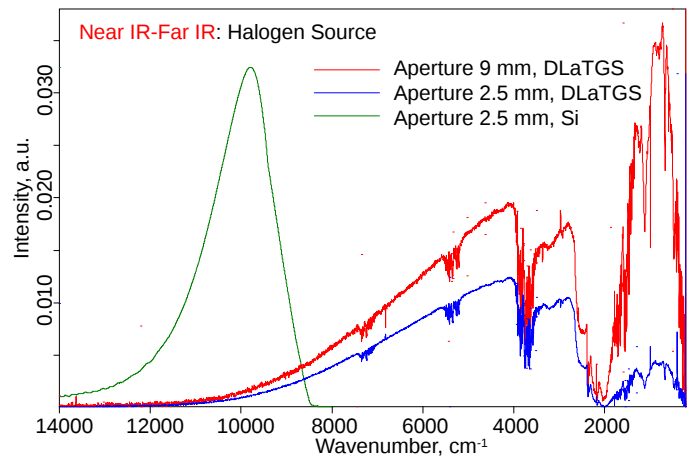}%
 \caption{\label{IR1} IR Spectral Coverage. Single beam spectra were obtained with Si and DLaTGS detectors, halogen light source and two different apertures. For DLaTGS measurements the spectrometer was not evacuated, therefore one can observe water vapor lines.}
 \end{figure}

\begin{figure}
 \includegraphics[width=0.5\textwidth]{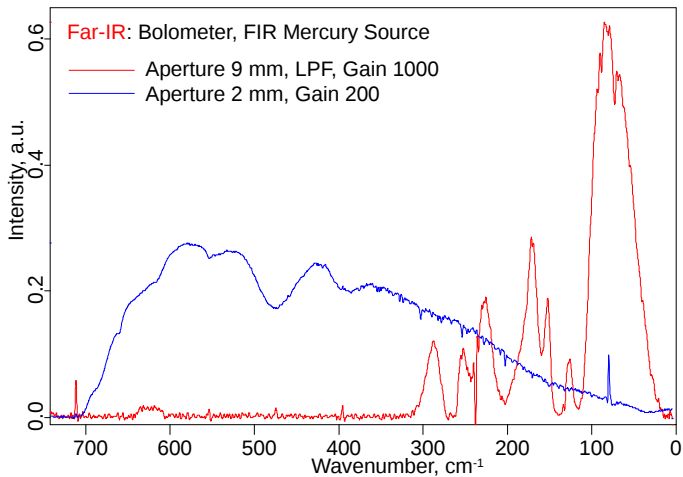}%
 \caption{\label{FIR} FIR Spectral Coverage. Single beam spectra were obtained with a bolometer detector and an FIR mercury light source. For measurements below 100~cm$^{-1}$ a low pass filter (LPF) was placed in front of the detector. In the latter case (red curve) absorptions are produced by the LPF.}
 \end{figure}

An overview of the spectral coverage associated with the upgraded spectrometer is presented in Fig.~\ref{IR1}. Using a Si detector (and a halogen source) one can almost reach the visible range. However, due to the imperfect planarity of the beam splitter, the overall efficiency of the Si detector decreases rapidly in going from the near-IR to the visible spectral region. This effect can also be seen when referring to the DLaTGS detector measurements. The beam splitter in the interferometer performs optimally at a wavelength $\lambda$ if its surfaces are optically flat, i.e. planarity is about $\lambda/20$. Otherwise phase error decreases the modulation intensity of the interferogram. This effect gives rise to different relative signal intensities in the mid-IR versus near-IR upon increasing  the aperture size. Broad absorption bands in diamond around 2000--2500~cm$^{-1}$ are two-phonon absorptions. These diamond absorptions could be reduced by using a thinner beam splitter with a smaller diameter. In the far-IR region it was possible to measure down to 15--30~cm$^{-1}$ with a mercury FIR lamp, Fig.~\ref{FIR}. 


\begin{figure}
 \includegraphics[width=0.5\textwidth]{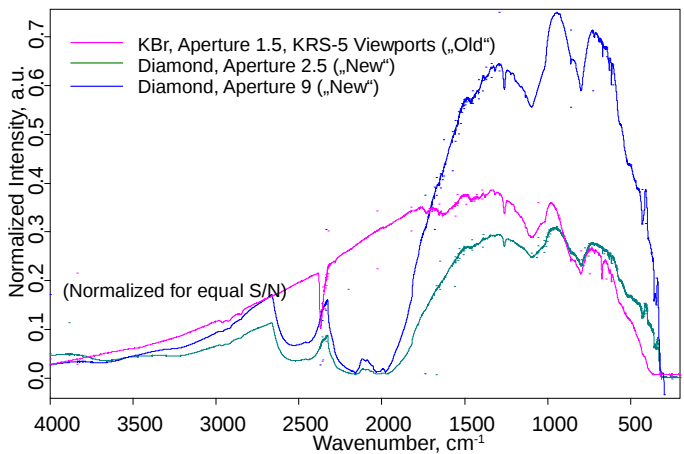}%
 \caption{\label{OldvsNew} KBr vs. diamond beam splitter in the Mid-IR. Single beam spectra were obtained with the CuGe detector and a Globar light source. They are normalized for an equal signal-to-noise ratio.}
 \end{figure}

Before the upgrade we had used a KBr beam splitter and wedged KRS-5 UHV windows in the UHV sample chamber. Therefore, we compare the performance of the new diamond optics configuration with measurements obtained with the old KBr/KRS-5 combination in Fig~\ref{OldvsNew}. We first note that,  outside of  the diamond phonon absorption band region, the overall signal intensity decreases  by roughly a factor of three   (using the same lamp and the same aperture). By increasing the aperture to 9~mm, one gets a stronger signal. At this aperture setting one might expect a decrease in resolution, but we do not observe it under our conditions (by measuring water vapor absorptions at a 0.25~cm$^{-1}$ resolution setting). A likely explanation for this is that the IR detector has a smaller active area than the focused light spot. 

\begin{figure}
 \includegraphics[width=0.5\textwidth]{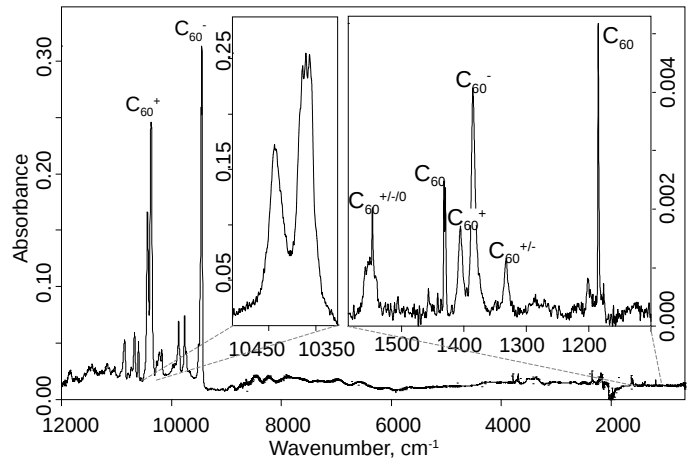}%
 \caption{\label{C60NIRMIR} Absorption spectra of C$_{60}^{+/-/0}$ isolated in neon matrix at 5~K measured with diamond optics and Si, MCT detectors together with Halogen / Globar light sources.}
 \end{figure}

\begin{figure}
 \includegraphics[width=0.5\textwidth]{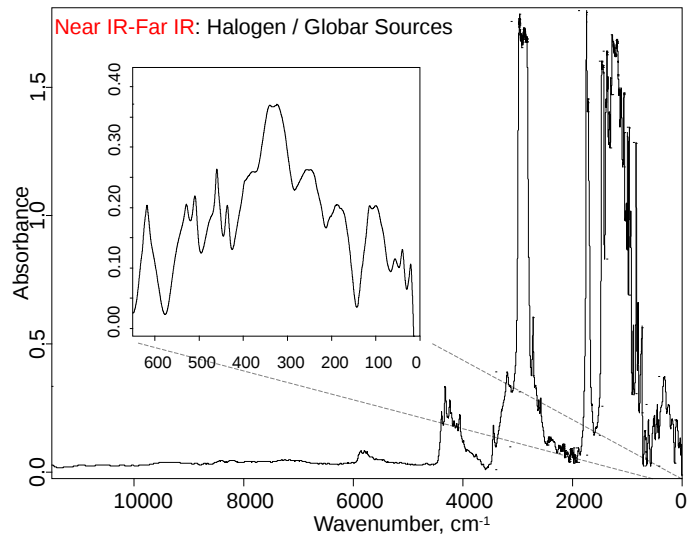}%
 \caption{\label{tesa} Absorption spectrum of one layer of Tesafilm\textsuperscript{\textregistered} at 300~K measured with Si, MCT and Bolometer detectors.}
 \end{figure}

An example of capability of the new setup for matrix isolation spectroscopy is shown in Fig.~\ref{C60NIRMIR}.  Here, the near-IR and mid-IR absorption spectra of fullerene ions were obtained with the diamond beam splitter. The sample contains about 0.4~nmol of C$_{60}^+$, 0.4~nmol of C$_{60}^-$ and 2~nmol of C$_{60}$ isolated in a cryogenic neon matrix. Measurements with the upgraded spectrometer save preparation time of a new sample, exchange of vacuum windows, bake-out time, and optical adjustment which in total can take one to three weeks. Further details about matrix isolation spectroscopy of fullerene ions can be found in our previous publications\cite{Depo2C60,Depo2C60dicat}. Another example of the performance achieved is demonstrated by the IR absorption spectrum of a piece of Tesafilm\textsuperscript{\textregistered}  adhesive tape, see Fig.~\ref{tesa}. This spectrum was recorded within an overall measurement time of 40 min, 10 min per detector (Si, MCT with Halogen source; MCT, Bolometer with Globar source). Additional examples of IR absorption spectra, measured with the upgraded FTIR spectrometer, can be found in the Supporting Information of our recent publication~\cite{Depo2C70}. 

\section{Summary}
We demonstrate that a synthetic diamond beam splitter without compensator plate can provide a significantly broader spectral coverage range (from 12000 to 15~cm$^{-1}$) than  present commercially available FTIR beam splitters. In combination with diamond windows this facilitates significantly faster acquisition of absorption spectra for matrix isolated ions. By increasing the planarity of the beam splitter it should become possible to work further into the visible range. Assuming suitable  light source / detector combinations are available, one could also measure further into the far-IR range. The two phonon absorption bands of diamond around 2000~cm$^{-1}$ could be reduced by using a thinner beam splitter. Since a beam splitter has practically no mechanical load, one might consider  producing and using  a 10-100~$\mu$m thin beam splitter, polished by a non-contact polishing method\cite{DiamPolish}. Using the diamond beam splitter the signal intensity in our system was roughly a factor of three less than obtained with a mid-IR KBr beam splitter. This could be compensated by using a larger aperture without significant loss of spectral resolution.

\begin{acknowledgments}
We thank Holger Halberstadt and Klaus Stree for building electronics, Dieter Waltz and the machine shop team for building the mechanical parts. We appreciate helpful discussions with G\"{u}nter Zachmann and Christian Riedinger (both Bruker Optik). We also thank them for providing electronic circuit diagrams of the HeNe laser preamplifier. This work was supported by the Deutsche Forschungsgemeinschaft (KA 972/10-1). We also acknowledge support by KIT and Land Baden-W\"{u}rttemberg. 
\end{acknowledgments}

\bibliography{depo2test}

\end{document}